\documentclass[aps,amssymb,showpacs,superscriptaddress,secnumarabic]{revtex4}
\usepackage{graphicx}
\usepackage{amsmath}
\numberwithin{equation}{section}
\usepackage{mathtools}
\usepackage{chngcntr}
\usepackage{bm}
\graphicspath{%
    {converted_graphics/}% inserted by PCTeX
    {/}% inserted by PCTeX
}
\begin{document}
\date{\today}
\title {Confined Random Walkers in Dimensions Higher Than One and\\ Analysis of Transmission of Infection in Epidemics}

\author{S. Sugaya} 
\email{satomi@unm.edu}
\author{V. M. Kenkre} 
\email{kenkre@unm.edu}
\affiliation{Consortium of the Americas for Interdisciplinary Science and the Department of Physics and Astronomy, University of New Mexico, Albuquerque, New Mexico 87131}

\begin{abstract}
A pair of random walkers, the motion of each of which in two dimensions is confined spatially by the action of a quadratic potential centered at different locations for the two walks, are analyzed in the context of reaction-diffusion. The application sought is to the process of transmission of infection in epidemics. The walkers are animals such as rodents in considerations of the Hantavirus epidemic, infected or susceptible, the reaction is the transmission of infection, and the confining potential represents the tendency of the animals to stay in the neighborhood of their home range centers. Calculations are based on a recently developed formalism (Kenkre and Sugaya, Bull. Math. Bio. 76, 3016 (2014)) structured around analytic solutions of a Smoluchowski equation and one of its aims is the resolution of peculiar but well-known problems of reaction-diffusion theory in 2-dimensions. In the present analysis, reaction occurs not at points but in spatial regions of dimensions larger than 0. The analysis uncovers interesting nonintuitive phenomena one of which is similar to that encountered in the 1-dimensional analysis given in the quoted article, and another specific to the fact that the reaction region is spatially extended. The analysis additionally provides a realistic description of observations on animals transmitting infection while moving on what is effectively a 2-dimensional landscape. Along with the general formalism and explicit 1-dimensional analysis given in 
Bull. Math. Bio. 76, 3016 (2014),
the present work forms a model calculational tool for the analysis for the transmission of infection in dilute systems.
\end{abstract}
. 

\maketitle

%05.40.-a	Fluctuation phenomena, random processes, noise, and Brownian motion (for fluctuations in superconductivity, see 74.40.-n; for statistical theory and fluctuations in nuclear reactions, see 24.60.-k; for fluctuations in plasma, see 52.25.Gj; for nonlinear dynamics and chaos, see 05.45.-a)
%82.40.Ck	Pattern formation in reactions with diffusion, flow and heat transfer (see also 47.54.-r Pattern selection; pattern formation and 47.32.C- Vortex dynamics in fluid dynamics)
%87.23.Cc	Population dynamics and ecological pattern formation
%82.20.-w	Chemical kinetics and dynamics
%05.40.Jc  	Brownian motion
%02.60.Cb	Numerical simulation; solution of equations
%82.20.Nk	Classical theories of reactions and/or energy transfer
%87.10.-e	General theory and mathematical aspects
%87.10.Ca	Analytical theories
%87.10.Ed	Ordinary differential equations (ODE), partial differential equations (PDE), 
%integrodifferential models
%87.19.rs 	Movement
%87.19.xd	Viral diseases
%%%%%%%%%%%%%%%%%%%%%%
\section{Background and Motivation for the Investigation}
%\begin{itemize}
%\item The problem of non-existent Laplace transform of $\nu(t)$ in 2-d or higher.
%\item 2-d is appropriate for Hantavirus.
%\item Original equation for point-infection.
%\item Summary of the paper.
%\end{itemize}

Random walks are ubiquitous in physics~\cite{wax} and biology~\cite{bbook}. Within the vast field of their investigation is the difficult and important area of \emph{confined} random walks. We present below analysis and practically usable results for random walkers moving under a confinement potential and interacting when they arrive within the influence of one another. We develop the theory as an extension of work we have published recently~\cite{KSBMB14} where we have provided a general formalism and illustrated it explicitly for a specific 1-dimensional case. The application in mind here, as in ref.~\cite{KSBMB14}, is to the description of transmission of epidemics such as the Hantavirus~\cite{Hanta}, wherein rodents moving randomly on the terrain pass on infection on encounter.

The reason for providing the 2-dimensional analysis in the present paper is two-fold. The first is that the  rodents involved in the spread of the Hantavirus do move on an effectively 2-dimensional (2-d) landscape and, although the detailed calculations in ref.~\cite{KSBMB14}, hereafter referred to as KS, have unearthed interesting results in a 1-dimensional (1-d) model, it is important to study whether the generalization to 2-d is straightforward or whether it changes significant features. There is, however, an additional reason for this study. In the continuum, the dimensionality of the region in which the reaction event occurs (in many instances trapping, in our present case the transmission of infection via contact) may be different from (smaller than) that in which the over-all motion of the walker(s) occurs. Thus, the analysis presented in KS had the over-all motion occurring on the 1-dimensional line but the encounter (reaction) occurred at a point, i.e., a zero-dimensional region. Reaction-diffusion theory in a spatial continuum of dimensions larger than 1 suffers from peculiar blow-ups if the reaction occurs \emph{at points}. Whereas our development in KS  was naturally built on point encounters on a line, motion in a plane such as in the terrain over which rodents execute their random walks can result in singular behavior for point encounters. An explicit resolution of the familiar blow-ups must be therefore implemented for \emph{realistic}  2-d and higher-dimensional systems.

The theory in the present paper has been constructed in two parts, one in which a straightforward extension of KS is presented but with both the reaction event and the diffusion occurring in higher-dimensional space, particularly 2-d as appropriate to rodents on the landscape. In the second part, we will display the blow-up problem for point encounters that fortuitously does not occur for 1-dimensional (over-all) motion but does in a plane and also show an explicit resolution of the problem by taking the reaction (passing of infection from one walker to another) to occur in an \emph{extended} region rather than at a point. In both parts we obtain usable expressions for the description of transmission of infection, recover a certain interesting non-monotonicity phenomenon  presented in the 1-d analysis of KS and, additionally, find here another phenomenon worthy of note that arises from the introduction of a new distance that characterizes the extent of the reaction region.

As explained in KS, an analytic theory of the transmission of infection is a challenging undertaking and is needed at this time because of recent developments in observations. Description of the transmission of infection in an epidemic derives its importance from human relevance as well as the theoretical difficulties of its intellectual challenge. Original contributions \cite{may,OkuboLevin,Hethcote,BAUER,DLM:2000} beginning with the study of Anderson and May \cite{may} involved concepts such as mass action, SIR, and the basic reproductive rate.  Spatial considerations were introduced into the investigations independently by various authors \cite{DLM:2000,OkuboLevin,AK:2002,AGUIRRE,CantrellCosner,KENKRE215:219,McKane,KENKRE203:233,McIinnisThesis}. Missing from all those studies were confinement features that arise in animal motion from the existence of home ranges. That these confinement considerations cannot be ignored became clear from the analysis of field observations \cite{GAKSMY:2005,jtb,McIinnisThesis}. Typically, the mean square displacement of the rodents extracted from mark-recapture experiments is found to saturate as time progresses rather than always growing linearly as unconfined random walks would predict. The saturation area represents the home ranges of the animals which, unmistakably, perform their random walks in a tethered manner, tending to return repeatedly to their shelters. Therefore, it became necessary to undertake a fundamental study of the transmission of infection in terms of interacting random walks specially under confinement.

Such a theory that takes explicit consideration of confinement has been developed \cite{KSBMB14} by the present authors in KS on the basis of pairwise interactions in a dilute system of random walkers (representing infected and susceptible animals) moving in $m$ dimensions. The model calculation is restricted to the time evolution of a representative pair of animals, one infected and one not, and is thus especially appropriate  to a dilute system of random walkers. The motion of each animal pair is taken to obey a Smoluchowski equation in 2$m$-dimensional space that combines diffusion with confinement of each animal to its particular home range. This passage to a space of twice the number of dimensions follows the formalism introduced by one of the present authors for the study of Frenkel exciton annihilation in molecular crystals \cite{vmk80}. The new feature of the present analysis (as that of KS) is the existence of home ranges, i.e., attractive centers for the walkers.  An additional (reaction) term that comes into play when the animals are in close proximity describes the process of infection. Analytic solutions are obtained, confirmed by numerical procedures, and shown \cite{KSBMB14} to predict a surprising effect of confinement \cite{SSK:2013}. The effect is that infection spread has a non-monotonic dependence on the diffusion constant and/or the extent of the attachment of the animals to the home ranges. Optimum values of these parameters exist for any given distance between the attractive centers. Any change from those values, involving faster/slower diffusion or shallower/steeper confinement, hinders the transmission of infection. A physical explanation has been provided in KS. Reduction to the simpler case of no home ranges has been demonstrated in full detail. Effective infection rates have been calculated, and it has been pointed out how one might use them in complex systems consisting of dense populations.

\section{Analysis in Higher Dimensions when the Dimensionality of Motion Equals that of Reaction }

Let the dimensionality of the over-all motion as well as that of the transmission of infection be the same and denoted by $m$. We denote the locations of the disease-carrying mouse, which we refer to as infected, and the one without but susceptible to the disease, which we refer to as susceptible, by $\bm{r}_{1}=(x_{1}^{1},x_{1}^{2},\dots,x_{1}^{m})$ and $\bm{r}_{2}=(x_{2}^{1},x_{2}^{2},\dots,x_{2}^{m})$, respectively, where the subscripts denote whether the mouse is infected or susceptible, and the superscripts denote the dimension. The probability density to find the walkers at their respective positions at time $t$ is denoted by $P(\bm{r}_{1},\bm{r}_{2},t)$. The mice are considered to be random walkers attracted to their home locations, $\bm{R}_{1}=(h_{1}^{1},h_{1}^{2},\dots,h_{1}^{m})$ and $\bm{R}_{2}=(h_{2}^{1},h_{2}^{2},\dots,h_{2}^{m})$, where $h_{i}^{j}$ is the home-coordinate in each dimension. 

\subsection{General Expression for the Infection Probability}
The equation of motion for the probability density is as in KS but now with an extended reaction region,
\begin{align}
\frac{\partial P(\bm{r}_{1},\bm{r}_{2},t)}{\partial t} &= \nabla_{1}\cdot\left[\gamma \left(\bm{r}_{1} - \bm{R}_{1}\right)P(\bm{r}_{1},\bm{r}_{2},t)\right]  + \nabla_{2}\cdot\left[\gamma \left(\bm{r}_{2} - \bm{R}_{2}\right)P(\bm{r}_{1},\bm{r}_{2},t)\right] 
 + D\left(\nabla_1^2+\nabla_2^2\right)P(\bm{r}_{1},\bm{r}_{2},t)\nonumber \\
 & - \mathcal{C}\int^{\prime}d\bm{r}^{\prime}_{1}d\bm{r}^{\prime}_{2}\,\delta(\bm{r}_{1}-\bm{r}^{\prime}_{1})\delta(\bm{r}_{2}-\bm{r}^{\prime}_{2})P(\bm{r}_{1},\bm{r}_{2},t),
 \label{start2}
\end{align}
where the primed integral is over the reaction region covered by $\bm{r}_{1}^{\prime}$ and $\bm{r}_{2}^{\prime}$. The third term on the right hand side describes the random motion of the random-walkers with $D$ being the diffusion constant. Their motion due to the attraction to their homes is described by the first two terms, where $\gamma$ is the strength of the attraction. Thus the walkers' motion is given by the Smoluchowski equation. Their infection-transmission interaction is given in the fourth term. The infection rate is denoted by $\mathcal{C}$ and has the unit $\left[1/\text{time}\right]$.\\

We focus on the homogeneous propagator, $\Pi(\bm{r}_{1},\bm{r}^{\prime}_{1},\bm{r}_{2},\bm{r}^{\prime}_{2},t)$, and obtain from it, the homogeneous solution,
\begin{equation}
\label{DefEta}
\eta(\bm{r}_{1},\bm{r}_{2},t) = \int\int d\bm{r}^{\prime}_{1}d\bm{r}^{\prime}_{2}\Pi(\bm{r}_{1},\bm{r}^{\prime}_{1},\bm{r}_{2},\bm{r}^{\prime}_{2},t)P(\bm{r}^{\prime}_{1},\bm{r}^{\prime}_{2},0), 
\end{equation}
with $P(\bm{r}^{\prime}_{1},\bm{r}^{\prime}_{2},0)$ being the initial condition, and the integral being over all space. This allows us to obtain the infection probability, $\mathcal{I}(t)$, the probability that the susceptible mouse is infected at time $t$, precisely as in KS,
\begin{equation}
\label{Itildemm1}
\mathcal{I}(t) = 1 - \int\int d\bm{r}_{1}d\bm{r}_{2}P(\bm{r}_{1},\bm{r}_{2},t),
\end{equation}
where the integral is over all space. With this definition, $\mathcal{I}(t)$ is obtained in the Laplace domain as
\begin{equation}
\label{Itildemm2}
\widetilde{\mathcal{I}}(\epsilon) = \frac{1}{\epsilon}\left[\frac{\widetilde{\mu}(\epsilon)}{1/\mathcal{C}+\widetilde{\nu}(\epsilon)}\right],
\end{equation}
where tilde denotes Laplace transform and $\epsilon$ is the Laplace variable. The functions $\mu(t)$ and $\nu(t)$ are given in terms of homogeneous quantities, $\eta(\bm{r}_{1},\bm{r}_{2},t)$ and $\Pi(\bm{r}_{1},\bm{r}^{\prime}_{1},\bm{r}_{2},\bm{r}^{\prime}_{2},t)$ by
\begin{equation}
\label{defMumm}
\mu(t) \equiv \int^{\prime}d\bm{r}_{1}\bm{r}_{2} \eta(\bm{r}_{1},\bm{r}_{2},t),
\end{equation}
\begin{equation}
\nu(t) \equiv \frac{\int^{\prime}d\bm{r}^{\prime}_{1}d\bm{r}^{\prime}_{2}\int^{\prime}\bm{r}_{1}d\bm{r}_{2}\Pi(\bm{r}_{1},\bm{x}^{\prime}_{1},\bm{r}_{2},\bm{r}^{\prime}_{2},t)}{\int^{\prime}d\bm{r}^{\prime}_{1}d\bm{r}^{\prime}_{2}}.
\end{equation}
Here, $\mu(t)$ is the probability to find the walkers within the infection region at time $t$ given their initial conditions, in the \emph{absence} of the infection phenomenon. Because of the extended reaction region, $\nu(t)$ here is obtained via the $\nu$-function method~\cite{vmknu}, where an ensemble average of the probability to find the walkers within the region at $t$ given that they were within the region initially, over the infection region. All these results are straight forward generalization of those in KS for arbitrary dimensions.

\subsection{Introduction of the Relative Coordinate}
\begin{figure}[h!] % float placement: (h)ere, page (t)op, page (b)ottom, other (p)age
  \centering
   \includegraphics[scale=0.7]{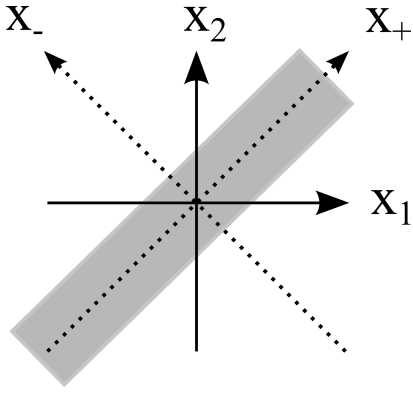}
   \caption{A 1-dimensional realization of the system showing the original coordinates, $x_1$-$x_2$ (solid lines). The CM-relative coordinates, $x_+$-$x_{-}$, are superimposed on the original axes in the dotted axes. The infection region is shown in gray, which lies along the CM ($x_+$) coordinate and is perpendicular to the relative ($x_-$) coordinate.}
   \label{fig:12TOpm}
\end{figure} 

A realistic scenario for an infection-transmission event is for it to occur when the mice come within a certain distance of one another. In such a case it is useful to make a coordinate transformation to the center of mass (CM) and relative coordinate system. To illustrate this explicitly, we refer the reader to Fig.~\ref{fig:12TOpm}. The 1-dimensional position phase-space in the original coordinate (labeled with $x_{1}^{1}\equiv x_{1}$ and $x_{2}^{1}\equiv x_{2}$) and CM-relative coordinate (labeled with $x_{+}$ and $x_{-}$) are superimposed. The gray region lying diagonal to the original coordinate indicates the infection region. Notice that this region lies parallel to the CM coordinate and perpendicular to the relative one, indicating that the transmission of infection occurs only dependent on the relative coordinate.\\

Thus, let the CM and the relative coordinates be denoted by $\bm{r}_{+}$ and $\bm{r}_{-}$, respectively. The specific transformation between $\bm{r}_{1}$, $\bm{r}_{2}$ and $\bm{r}_{+}$, $\bm{r}_{-}$ is given by
\begin{equation}
\label{12toPM}
\bm{r}_{\pm}=\frac{1}{\sqrt{2}}\left(\bm{r}_{1}\pm \bm{r}_{2}\right) \quad \Longleftrightarrow \quad \bm{r}_{1,2} = \frac{1}{\sqrt{2}}\left(\bm{r}_{+}\pm \bm{r}_{-}\right),
\end{equation}
where the factor of $1/\sqrt{2}$ is chosen to preserve the form of the motion part of the equation. With this transformation, the equation of motion becomes
\begin{align}
\frac{\partial P(\bm{r}_{+},\bm{r}_{-},t)}{\partial t} &= \nabla_{+}\cdot\left[\gamma \left(\bm{r}_{+} - \bm{R}_{+}\right)P(\bm{r}_{+},\bm{r}_{-},t)\right]  + \nabla_{-}\cdot\left[\gamma \left(\bm{r}_{-} - \bm{R}_{-}\right)P(\bm{r}_{+},\bm{r}_{-},t)\right] 
 + D\left(\nabla_{+}^2+\nabla_{-}^2\right)P(\bm{r}_{+},\bm{r}_{-},t)\nonumber \\
 &- \mathcal{C}\int^{\prime}d\bm{x}^{\prime}_{+}d\bm{x}^{\prime}_{-}\,\delta(\bm{x}_{+}-\bm{x}^{\prime}_{+})\delta(\bm{x}_{-}-\bm{x}^{\prime}_{-})P(\bm{x}_{+},\bm{x}_{-},t),
\end{align}
where $\bm{R}_{\pm}$ are given by 
\begin{equation}
\label{12toPMh}
\bm{R}_{\pm} = \frac{1}{\sqrt{2}}\left(\bm{R}_{1}\pm \bm{R}_{2}\right).
\end{equation} 
%We note that the coordinate transformation rule for a delta-function is given by
%\begin{equation}
%\delta(\bm{a}) = \frac{1}{|J|}\delta(\bm{b}),
%\end{equation}
%where the transformation between an arbitrary coordinate vectors $\bm{a}$ and $\bm{b}$ is assumed not singular, and $J$ is the Jacobian of the transformation, which turns out to be $-1$ between each element of $\bm{r}_{1}$ and $\bm{r}_{2}$ here. \\

Following the same procedure as procedure in the system of CM-relative coordinates we find 
\begin{equation}
\label{nuPMmm}
\nu(t)\equiv \frac{\int^{\prime}d\bm{r}^{\prime}_{+}d\bm{r}^{\prime}_{-}\int^{\prime}d\bm{r}_{+}d\bm{r}_{-}\,\Pi(\bm{r}_{+},\bm{r}^{\prime}_{+},\bm{r}^{\prime}_{-},\bm{r}^{\prime}_{-}t)}{\int^{\prime}d\bm{r}^{\prime}_{+}d\bm{r}^{\prime}_{-}},
\end{equation}
and
\begin{equation}
\mu(t) \equiv \int^{\prime}d\bm{r}_{+}d\bm{r}_{-}\,\eta(\bm{r}_{+},\bm{r}_{-},t).
\end{equation} 

The propagator in this coordinate, $\Pi(\bm{r}_{+},\bm{r}^{\prime}_{+},\bm{r}_{-},\bm{r}^{\prime}_{-},t)$, is given by
\begin{equation}
\label{PropPMsD}
\Pi(\bm{r}_{+},\bm{r}_{+}^{\prime},\bm{r}_{-},\bm{r}_{-}^{\prime},t) = \left(\frac{1}{\sqrt{4\pi D \mathcal{T}(t)}}\right)^{2m}\,\prod_{\beta=1}^{m}e^{-\frac{\left(x_{+}^{\beta}-h_{+}^{\beta}-(x_{+}^{\prime\beta}-h_{+}^{\beta})e^{-\gamma t}\right)^2+\left(x_{-}^{\beta}-h_{-}^{\beta}-(x_{-}^{\prime\beta}-h_{-}^{\beta})e^{-\gamma t}\right)^2}{4D\mathcal{T}(t)}}.
\end{equation}
Since the entire CM coordinate is part of the infection region as explained above, the primed integral is appropriately written as
\begin{equation}
\int^{\prime}d\bm{r}_{+}d\bm{r}_{-} \rightarrow \int_{\text{all space}}d\bm{r}_{+}\int^{\prime}d\bm{r}_{-}. 
\end{equation}
Using this relation and performing the first integral in the numerator of the definition of $\nu(t)$ given in Eq.~(\ref{nuPMmm}) results in 
\begin{align}
&\int_{\text{all space}}d\bm{r}_{+}\int^{\prime}d\bm{r}_{-}\,\Pi(\bm{r}_{+},\bm{r}_{+}^{\prime},\bm{r}_{-},\bm{r}_{-}^{\prime},t) \nonumber \\
& =\int_{\text{all space}}d\bm{r}_{+}\int^{\prime}d\bm{r}_{-}\left(\frac{1}{\sqrt{4\pi D \mathcal{T}(t)}}\right)^{2m}\,\prod_{\beta=1}^{2m}e^{-\frac{\left(x_{+}^{\beta}-h_{+}^{\beta}-(x_{+}^{\prime\beta}-h_{+}^{\beta})e^{-\gamma t}\right)^2+\left(x_{-}^{\beta}-h_{-}^{\beta}-(x_{-}^{\prime\beta}-h_{-}^{\beta})e^{-\gamma t}\right)^2}{4D\mathcal{T}(t)}}\nonumber \\
& =\int^{\prime}d\bm{r}_{-}\left(\frac{1}{\sqrt{4\pi D \mathcal{T}(t)}}\right)^m\,\prod_{\beta=1}^{m}e^{-\frac{\left(x_{-}^{\beta}-h_{-}^{\beta}-(x_{-}^{\prime\beta}-h_{-}^{\beta})e^{-\gamma t}\right)^2}{4D\mathcal{T}(t)}}. \label{pirelmm}
\end{align}
Since we see clearly that the CM coordinate is integrated entirely out of the problem, we need focus only on the relative coordinate. Let us define the integrand of Eq.~(\ref{pirelmm}) as
\begin{equation}
\label{pirel}
\Pi(\bm{r}_{-},\bm{r}^{\prime}_{-},t) \equiv \left(\frac{1}{\sqrt{4\pi D \mathcal{T}(t)}}\right)^m\,\prod_{\beta=1}^{m}e^{-\frac{\left(x_{-}^{\beta}-h_{-}^{\beta}-(x_{-}^{\prime\beta}-h_{-}^{\beta})e^{-\gamma t}\right)^2}{4D\mathcal{T}(t)}}.
\end{equation}
With this result, the $\nu$-function can be written as
\begin{equation}
\nu(t) = \frac{\int_{\text{all space}}d\bm{r}^{\prime}_{+}\int^{\prime}d\bm{r}^{\prime}_{-}\int^{\prime}d\bm{r}_{-}\,\Pi(\bm{r}_{-},\bm{r}^{\prime}_{-},t)}{\int_{\text{all space}}d\bm{r}^{\prime}_{+}\int^{\prime}d\bm{r}^{\prime}_{-}}.
\end{equation}
The propagator does not depend on $\bm{r}^{\prime}_{+}$. The integrals over $\bm{r}^{\prime}_{+}$ in the numerator and the denominator cancel, leading to
\begin{equation}
\label{mainPMnumm}
\nu(t) = \frac{\int^{\prime}d\bm{r}^{\prime}_{-}\int^{\prime}d\bm{r}_{-}\,\Pi(\bm{r}_{-},\bm{r}^{\prime}_{-},t)}{\int^{\prime}d\bm{r}^{\prime}_{-}}.
\end{equation}

Let us consider a delta-function initial condition of the walkers for the rest of this paper, \emph{i.e.}, $P(\bm{r}_{1},\bm{r}_{2},0) = \delta\left(\bm{r}_{1}-\bm{r}^{0}_{1}\right)\delta\left(\bm{r}_{2}-\bm{r}^{0}_{2}\right)$, which translates to $P(\bm{r}_{+},\bm{r}_{-},0) = \delta\left(\bm{r}_{+}-\bm{r}^{0}_{+}\right)\delta\left(\bm{r}_{-}-\bm{r}^{0}_{-}\right)$ according to Eq.~(\ref{12toPM}). Therefore $\mu(t)$ becomes
\begin{equation}
\label{mainPMmumm}
\mu(t) = \int^{\prime}d\bm{r}_{-}\,\Pi(\bm{r}_{-},\bm{r}^{0}_{-},t).
\end{equation}
Most actual situations correspond to a 2-dimensional space because the mice that transmit the infection move about on the open terrain which can be considered flat for all practical purposes. Therefore, we give further analysis in 2-dimensions.\\

\subsection{Reaction Region as a Disk in the Plane}
For the ease of notation, we let the relative coordinate in the first dimension be denoted by $x^{1}_{-}\equiv x$, that of the second dimension by $x_{-}^{2}\equiv y$, the initial condition by $x_{-}^{0,1}\equiv x^{0}$ and $x_{-}^{0,2}\equiv y^{0}$, the relative home-center coordinate in the first dimension by $h_{-}^{1} \equiv h$, and that of the second dimension by $h_{-}^{2}\equiv f$. The propagator corresponding to Eq.~(\ref{pirel}) then becomes
\begin{equation}
\Pi(x,x',y,y',t) = \left(\frac{1}{4\pi D \mathcal{T}(t)}\right)\,e^{-\frac{\left(x-h-(x'-h)e^{-\gamma t}\right)^2+\left(y-f-(y'-f)e^{-\gamma t}\right)^2}{4D\mathcal{T}(t)}},
\end{equation}
and $\nu(t)$ and $\mu(t)$ are given by
\begin{equation}
\nu(t) = \frac{\int^{\prime}dx'dy'\int^{\prime}dxdy\,\Pi(x,x',y,y',t)}{\int^{\prime}dx'dy'},
\end{equation}
and
\begin{equation}
\mu(t) = \int^{\prime}dxdy\Pi(x,x^{0},y,y^{0},t).
\end{equation}
Suppose that the infection is transmitted to the susceptible mouse at a rate $\mathcal{C}$ when the mice come within a distance $b$ of one another. This means that the infection region is described by $x^2+y^2\le b^2$. Hence it is clear that the primed integral over the infection region is expressed simpler in the corresponding polar coordinate, where it becomes a circular disk of radius $b$, \emph{i.e.},
\begin{equation}
\int^{\prime}dxdy= \int_{0}^{b}dr\int_{0}^{2\pi}rd\phi.
\end{equation}
Since the polar variables, $r$ and $\phi$, are given by
\begin{equation}
\label{RelToPolar}
\left\lbrace
\begin{tabular}{l}
	$r^2 = x^{2}+y^{2}$\\
	$\phi = \tan^{-1} (y/x)$
\end{tabular}
\right.
\Longleftrightarrow
\left\lbrace
\begin{tabular}{l}
	$x = r \cos\phi$\\
	$y = r \sin\phi_{\quad \quad .}$
\end{tabular}
\right. ,
\end{equation}
the relative-home-center coordinates transformed as
\begin{equation}
\label{RelToPolar}
\left\lbrace
\begin{tabular}{l}
	$H^2 = h^2+f^2$\\
	$\omega = \tan^{-1}(f/h)$
\end{tabular}
\right.
\Longleftrightarrow
\left\lbrace
\begin{tabular}{l}
	$h = H \cos\omega$\\
	$f = H \sin\omega_{\quad \quad .}$
\end{tabular}
\right. 
\end{equation}
In these polar coordinates, the propagator takes the form
\begin{equation}
\label{PiPolar1}
\Pi(r,r',\phi,\phi',t) =\frac{1}{4\pi D\mathcal{T}(t)}e^{-\frac{r^2+\mathcal{F}^2(r',\phi',t)-2r\mathcal{F}(r',\phi',t)\cos(\phi-\Phi(r',\phi',t))}{4D\mathcal{T}(t)}}.
\end{equation}
Here, the functions $\mathcal{F}$ and $\Omega$ are given in terms of $\gamma$, $H$, and $\mathcal{T}(t)$ as
\begin{equation}
\mathcal{F}^2(r',\phi',t) = r^{\prime 2} -2\gamma\left[r^{\prime 2}+H^2-2r'H\cos(\phi'-\omega)\right]\mathcal{T}(t),
\end{equation}
\begin{equation}
\Omega(r',\phi',t) = \tan^{-1}\left(\frac{(r'\sin\phi'-H\sin\omega)e^{-\gamma t}+H\sin\omega}{(r'\cos\phi'-H\cos\omega)e^{-\gamma t}+H\cos\omega}\right).
\end{equation}
In computing the important quantity $\nu(t)$ is given by
\begin{equation}
\nu(t) = \frac{\int_{0}^{b}dr'\int_{0}^{2\pi}r'd\phi'\int_{0}^{b}dr\int_{0}^{2\pi}rd\phi\,\Pi(r,r',\phi,\phi',t)}{\int_{0}^{b}dr'\int_{0}^{2\pi}r'd\phi'}
\end{equation}
requires the calculation of the integral over $\phi$ of $\Pi(r,r'\phi,\phi',t)$. The analytic evaluation of this integral is, which is nothing other than the propagator to find the walkers somewhere within the disk, being separated by a distance $r\le b$ at $t$, given that they were separated by $r'\le b$ and angle $\phi'$ at the beginning, straightforward. Calling this integral as $\Pi(r,r',\phi',t)$, we see that
\begin{align}
\label{PiPolar2}
\Pi(r,r',\phi',t) & \equiv \int_{0}^{2\pi}d\phi\,\Pi(r,r',\phi,\phi',t) \\
& = \frac{1}{4\pi D \mathcal{T}(t)}e^{-\frac{r^2+\mathcal{F}^2(r',\phi',t)}{4D\mathcal{T}(t)}}\int_{0}^{2\pi}d\phi\,e^{\frac{2r\mathcal{F}(r',\phi',t)\cos(\phi-\Phi(r',\phi',t))}{4D\mathcal{T}(t)}}\\
& =\frac{1}{2 D \mathcal{T}(t)}e^{-\frac{r^2+\mathcal{F}^2(r',\phi',t)}{4D\mathcal{T}(t)}}\text{I}_{0}\left(\frac{r\mathcal{F}(r',\phi',t)}{2D\mathcal{T}}\right),
\end{align}
where I$_{0}(z)$ is the 0-th order modified Bessel function of the first kind. This expression is similar to and a generalization of the expression obtained recently by Spendier and Kenkre~\cite{KSinJPC}, and earlier in the context of heat conduction by Carslaw and Jaeger~\cite{CARSLAW}. Then, in terms of $\Pi(r,r',\phi',t)$, $\nu(t)$ and $\mu(t)$ are given by
\begin{equation}
\label{nuDisk}
\nu(t) = \frac{1}{\pi b^2}\int_{0}^{b}dr'\int_{0}^{2\pi}d\phi'\int_{0}^{b}dr\,\,rr'\Pi(r,r',\phi',t),
\end{equation}
\begin{equation}
\label{muDisk}
\mu(t) =\int_{0}^{b}dr\,\,r\Pi(r,r^{0},\phi^{0},t).
\end{equation}

We now examine the short-time behavior of $\nu(t)$ by taking limit $t\rightarrow0$ in Eq.~(\ref{nuDisk}). When $t\rightarrow 0$, the functions $\mathcal{T}(t)$ and $\mathcal{F}(r',\phi',t)$ become
\begin{align}
& \mathcal{T}(t\rightarrow0) \rightarrow t \\
& \mathcal{F}^2(r',\phi',t\rightarrow0)\rightarrow r^{\prime 2},
\end{align}
which in tern makes $\Pi(r,r',\phi',t)$ to become
\begin{equation}
\Pi(r,r',\phi',t\rightarrow 0) \rightarrow \frac{1}{2Dt}e^{-\frac{r^2+r^{\prime 2}}{2Dt}}\text{I}_{0}\left(\frac{rr'}{2Dt}\right).
\end{equation}
Furthermore, the argument of the Bessel function diverges and it is appropriate to express I$_{0}(z)$ in terms of its asymptotic form~\cite{Abramowitz:1970}, which is
\begin{equation}
\label{I0tShort}
\text{I}_{0}(z) \sim \frac{1}{\sqrt{2\pi z}}e^{z}, \quad \text{for large $z$.}
\end{equation}
With this, $\Pi(r,r',\phi',t\rightarrow0)$ becomes
\begin{equation}
\Pi(r,r',\phi',t\rightarrow0) \rightarrow \frac{1}{\sqrt{rr'}}\frac{1}{\sqrt{4\pi Dt}}e^{-\frac{(r-r')^2}{4Dt}} \rightarrow \frac{1}{\sqrt{rr'}}\delta(r-r').
\end{equation}
Using this result, $\nu(t\rightarrow0)$ is 
\begin{equation}
\nu(t\rightarrow0) \rightarrow \frac{1}{\pi b^2}\int_{0}^{b}dr'\int_{0}^{2\pi}d\phi'\int_{0}^{b}dr\,\,rr'\sqrt{\frac{1}{rr'}}\delta(r-r')= \frac{1}{\pi b^2}\int_{0}^{b}dr'\int_{0}^{2\pi}d\phi'\int_{0}^{b}dr\,\,r' = 1.
\end{equation}
We thus see that $\nu(t)$ is well behaved at short times and its Laplace transform does exist.\\

**Nitant to tighten the paragraph below in the next path**
Further calculations to arrive at the infection probability $\mathcal{I}(t)$ requires numerical means: first, $\nu(t)$ and $\mu(t)$ given in Eq's.~(\ref{nuDisk}) and (\ref{muDisk}) are calculated by numerically evaluating the integrals. Then they are numerically Laplace transformed, and the result is used to obtain the infection probability in the Laplace domain, $\widetilde{\mathcal{I}}(\epsilon)$, through Eq.~(\ref{Itildemm2}). Finally, this $\widetilde{\mathcal{I}}(\epsilon)$ is numerically inverted to obtain $\mathcal{I}(t)$.\\

\subsection{Non-monotonicity in the Dependence of the Strength of the Confinement}
\begin{figure}[h!] % float placement: (h)ere, page (t)op, page (b)ottom, other (p)age
  \centering
   \includegraphics[width= 0.8\columnwidth]{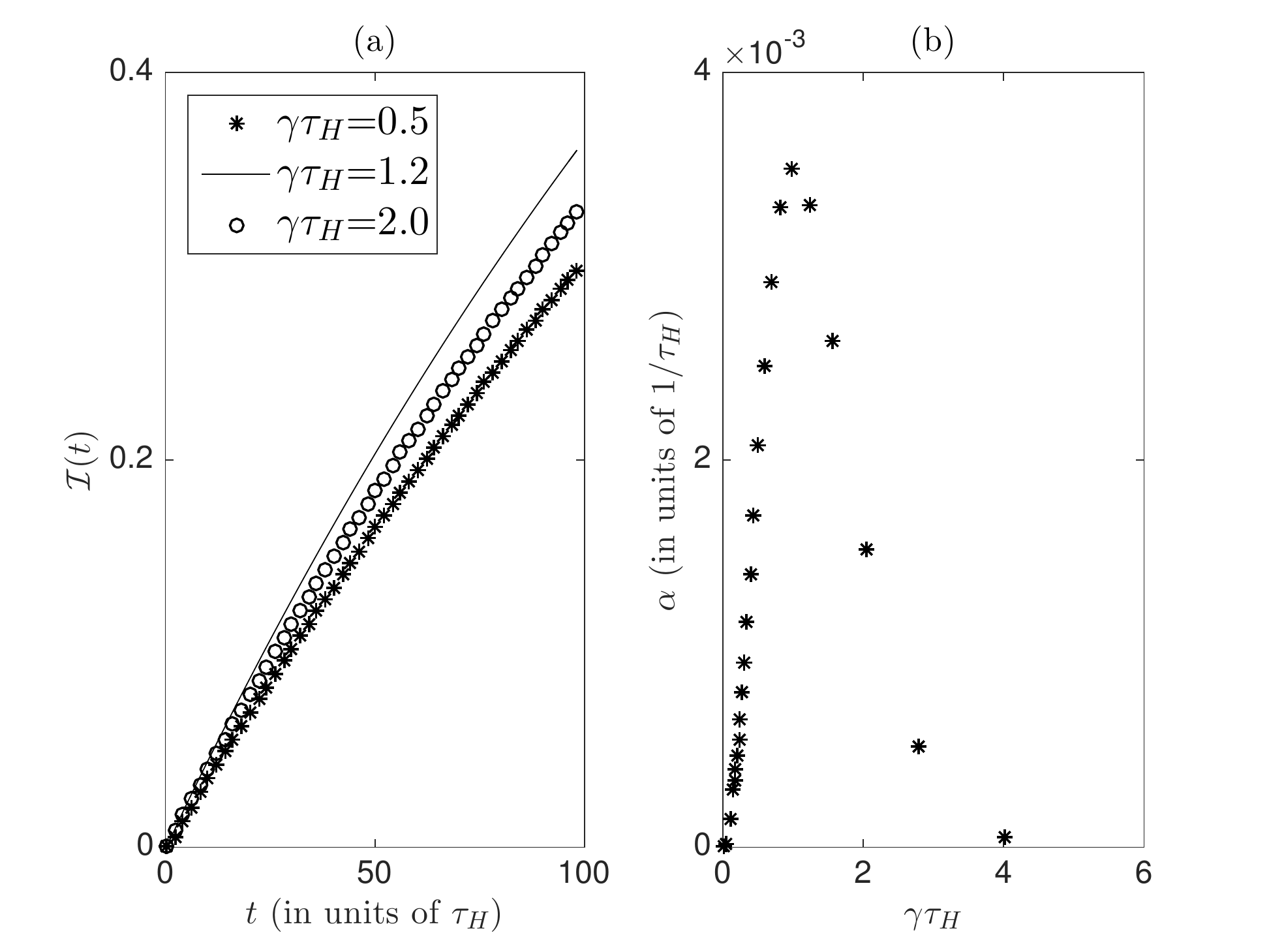}
   \caption{(a)The non-monotonic effect of infection probability as a function of increasing $\gamma\tau_H$. The latter dimensionless parameter characterizes the strength of confinement. The infection probability $\mathcal{I}(t)$ is plotted against time in units of the diffusive time, $\tau_H$, for the walkers to traverse the inter-home distance $H$. Infection curves are produced for the values of $\gamma\tau_H$ = 0.5, 1.2, and 2.0. (b) The effective rate $\alpha$ scaled to $1/\tau_H$ plotted against $\gamma\tau_H$, whose peak indicates the optimal value of $\gamma\tau_H$ for infection. The rate of infection $\mathcal{C}$ is kept constant at 0.05 in units of $1/\tau_H$.}
   \label{fig:SK_Disk2D}
\end{figure} 
The infection probability, obtained numerically, shows the non-monotonic effect of the home ranges of the walkers. On panel (a) of Fig.~\ref{fig:SK_Disk2D}, the infection probability against time (scaled to the diffusive time, $\tau_H = H^2/2D$) is given. As the magnitude of the confinement of the walkers to their respective homes is increased from $\gamma\tau_H$ = 0.5 (asterisk), 1.2 (solid line), to 2.0 (circles), the infection curve behaves non-monotonically, as observed in the 1-dimensional result given in KS.\\

In addition to $\mathcal{I}(t)$, as in KS, the effective rate of infection, $\alpha$, can be defined to efficiently characterize the effect of the confining potentials on $\mathcal{I}(t)$. In order to find such a quantity, we asked: what would the rate $\alpha$ be in terms of $\nu(t)$ and $\mu(t)$, if $\mathcal{I}(t)$ were a decaying exponential of the form $\mathcal{I}(t)=1-e^{-\alpha t}\,$? In the Laplace domain, such $\mathcal{I}(t)$ is given by $\widetilde{\mathcal{I}}(\epsilon)=\alpha/\left[\epsilon(\epsilon+\alpha)\right]$, which is equated to the right hand side of Eq.~(\ref{Itildemm2}) to be solved for $\epsilon$-dependent quantity, $\widetilde{\alpha}(\epsilon)$. Finally, Abelian theorem was applied to define the effective rate: $\alpha\equiv lim_{\epsilon\rightarrow 0}\,\left[ \epsilon \widetilde{\alpha}(\epsilon)\right]$, which results in
\begin{equation}
\label{alphaDefmm}
\alpha = \frac{\mu(\infty)}{1/\mathcal{C}+1/\mathcal{M}},
\end{equation}
where $\mathcal{M}$, which we call the motion parameter is defined as
\begin{equation}
\frac{1}{\mathcal{M}} = \int_{0}^{\infty}dt\,\left[\nu(t)-\mu(t)\right].
\end{equation} 
In panel (b) of Fig.~\ref{fig:SK_Disk2D}, $\alpha$ scaled to $1/\tau_H$ is plotted against $\gamma\tau_H$. The peaking behavior of $\alpha$ as a function of $\gamma\tau_H$ shows that the degree of the confinement of the random-walkers to their respective homes affect the infection probability non-monotonically.\\

This result reconfirms the generalization of the effect of confinement found in 1-dimension in KS. The existence of the home ranges of the walkers makes their probability density to become Gaussians in the steady state. As seen in the expressions of $\tilde{\mathcal{I}}(\epsilon)$ in Eq.~(\ref{Itildemm2}) and $\alpha$ in Eq.~(\ref{alphaDefmm}), the effects of confinement enter these quantities through the homogeneous probabilities, $\mu(t)$ and $\nu(t)$ to find the walkers within the infection region at time $t$, given their initial conditions, and given that they were within the region initially, respectively. The probability density to find the walkers together, in the steady state, varies non-monotonically as a function of the confinement strength, and thus gives rise to the non-monotonic effect in $\mathcal{I}(t)$ and $\alpha$.\\

We now discuss a more general situation, where the dimensionality of the infection region and that of the motion of walkers differ. \\

\section{Analysis in Higher Dimensions when the Dimensionality of Motion Exceeds that of Reaction}
Although the above analysis is based on the assumption that the dimensionality of the walker equals that of the infection region, practical situations exist in which this is not true. A clear example is a pair of walkers each executing a 1-d random walk and transmitting infection on encounter at a point. This example indeed underlies the analysis of ref.~\cite{KSBMB14}. A mathematical problem arises if each of the walkers executes a 2-d random walk but the infection region is still a point. Singularities appear in the short-time behavior of the $\nu$-function causing its Laplace transform impossible to be evaluated.\\

Generally, for a point infection-region in $m$-dimensions, $\nu(t)$ is given by~\cite{KSBMB14}
\begin{align}
\label{nuColocate}
\nu(t) &= \frac{\int_{\text{all space}}d^{m}\bm{r}^{\prime}_1\int_{\text{all space}}d^{m}\bm{r}_1\,\,\Pi(\bm{r}_1,\bm{r}'_1,\bm{r}_1,\bm{r}'_1,t)}{\int_{\text{all space}}d^{m}\bm{r}^{\prime}_1}\\ 
& = \left(\frac{1}{\sqrt{8\pi D\mathcal{T}(t)}}\right)^{m}\prod_{\beta=1}^{m}e^{-\frac{\left[\left(h_1^{\beta}-h_2^{\beta}\right)\left(1-e^{-\gamma t}\right)\right]^2}{8D\mathcal{T}(t)}}.
\end{align}
In the limit $t\rightarrow 0$, this $\nu$-function behaves as
\begin{equation}
\nu(t\rightarrow 0 )\rightarrow t^{-m/2}
\end{equation}
to the lowest order of $t$. Because of this from, the Laplace transform of $\nu(t)$ exists only when $m=1$, and does not exist for $m>1$. What this means is that while 1-d walkers transmitting infection on meeting at a point do not raise any problems for the theory as developed in KS, an extension of the theory to 2-d walkers doing the same presents a singularity problem.\\

We show how this problem can be solved by expanding the dimensionality of the infection region, and give an example in 2-dimensions.\\

\subsection{Resolution of Point Encounter Problems by Expansion of the Reaction Region}
Consider a situation where infection transmission occurs when the walkers are within a certain region of dimensionality $\theta$ that is smaller than that of the motion: $\theta< 2m$. For such a case, we write, for the equation of motion for $P(\bm{r}_{1},\bm{r}_{2},t)$,
\begin{align}
\frac{\partial P(\bm{r}_{1},\bm{r}_{2},t)}{\partial t} &= \nabla_{1}\cdot\left[\gamma \left(\bm{r}_{1} - \bm{R}_{1}\right)P(\bm{r}_{1},\bm{r}_{2},t)\right]  + \nabla_{2}\cdot\left[\gamma \left(\bm{r}_{2} - \bm{R}_{2}\right)P(\bm{r}_{1},\bm{r}_{2},t)\right] 
 + D\left(\nabla_1^2+\nabla_2^2\right)P(\bm{r}_{1},\bm{r}_{2},t)\nonumber \\
 & - \mathcal{C}_{(2m-\theta)}\int^{\prime}d^{\theta_{1}}\bm{r}^{\prime}_{1}d^{\theta_{2}}\bm{r}^{\prime}_{2}\,\delta^{m}(\bm{r}_{1}-\bm{r}^{\prime}_{1})\delta^{m}(\bm{r}_{2}-\bm{r}^{\prime}_{2})P(\bm{r}_{1},\bm{r}_{2},t).
 \label{start2}
\end{align}
The first three terms on the right hand side describe the motion of the random-walkers precisely as in Eq.~(\ref{start2}). The departure from Eq.~(\ref{start2}) is seen in the fourth term, in the dimensionality of the differential elements, $d^{\theta_{1}}\bm{r}_{1}$ and $d^{\theta_{2}}\bm{r}_{2}$, and the infection rate $\mathcal{C}_{(2m-\theta)}$. The dimensionalities of the infection region for infected and susceptible mice are given by $\theta_{1}$ and $\theta_{2}$, respectively, where $\theta=\theta_{1}+\theta_{2}$. The infection rate is denoted by $\mathcal{C}_{(2m-\theta)}$ to reflects its dimensionality:$\left[\text{length}^{(2m-\theta)}/\text{time}\right]$.\\

We follow the previous arguments to arrive at the expression of the infection probability in the Laplace Domain. In terms of the propagator, $\Pi(\bm{r}_{1},\bm{r}^{\prime}_{1},\bm{r}_{2},\bm{r}^{\prime}_{2},t)$, the solution in the Laplace space is given by
\begin{equation}
\label{PseudoSolutmt}
\widetilde{P}(\bm{r}_{1},\bm{r}_{2},\epsilon) = \widetilde{\eta}(\bm{r}_{1},\bm{r}_{2},\epsilon) - \mathcal{C}_{(2m-\theta)}\int^{\prime}d^{\theta_{1}}\bm{r}^{\prime}_{1}d^{\theta_{2}}\bm{r}^{\prime}_{2}\,\widetilde{P}(\bm{r}^{\prime}_{1},\bm{r}^{\prime}_{2},\epsilon)\widetilde{\Pi}(\bm{r}_{1},\bm{x}^{\prime}_{1},\bm{r}_{2},\bm{r}^{\prime}_{2},\epsilon),
\end{equation}
where $\eta(\bm{r}_{1},\bm{r}_{2},t)$ is the homogeneous solution given in Eq.~(\ref{DefEta}). The definition of the infection probability (Eq.~(\ref{Itildemm1})) yields
\begin{equation}
\label{Itildemt1}
\widetilde{\mathcal{I}}(\epsilon) = \frac{\mathcal{C}_{(2m-\theta)}}{\epsilon}\int^{\prime}d^{\theta_{1}}\bm{r}^{\prime}_{1}d^{\theta_{2}}\bm{r}^{\prime}_{2}\,\widetilde{P}(\bm{r}^{\prime}_{1},\bm{r}^{\prime}_{2},\epsilon).
\end{equation} 
Applying the defect technique~\cite{montroll} to Eq.~(\ref{PseudoSolutmt}), we integrate in the variables $\bm{r}_{1}$ and $\bm{r}_{2}$ over the infection region, \emph{i.e.},
\begin{align}
&\int^{\prime}d^{\theta_{1}}\bm{r}_{1}d^{\theta_{2}}\bm{r}_{2}\widetilde{P}(\bm{r}_{1},\bm{r}_{2},\epsilon) =\int^{\prime}d^{\theta_{1}}\bm{r}_{1}d^{\theta_{2}}\bm{r}_{2}\widetilde{\eta}(\bm{r}_{1},\bm{r}_{2},\epsilon)\nonumber\\
&-\mathcal{C}_{(2m-\theta)}\int^{\prime}d^{\theta_{1}}\bm{r}^{\prime}_{1}d^{\theta_{2}}\bm{r}^{\prime}_{2}\widetilde{P}(\bm{r}^{\prime}_{1},\bm{r}^{\prime}_{2},\epsilon)\int^{\prime}d^{\theta_{1}}\bm{r}_{1}d^{\theta_{2}}\bm{r}_{2}\widetilde{\Pi}(\bm{r}_{1},\bm{r}^{\prime}_{1},\bm{r}_{2},\bm{r}^{\prime}_{2}\epsilon). \label{defectmt}
\end{align}
We define $\nu_{(2m-\theta)}(t)$ by the use of the $\nu$-function method~\cite{vmknu}:
\begin{equation}
\nu_{(2m-\theta)}(t)\equiv \frac{\int^{\prime}d^{\theta_{1}}\bm{r}^{\prime}_{1}d^{\theta_{2}}\bm{r}^{\prime}_{2}\int^{\prime}d^{\theta_{1}}\bm{r}_{1}d^{\theta_{2}}\bm{r}_{2}\Pi(\bm{r}_{1},\bm{r}^{\prime}_{1},\bm{r}_{2},\bm{r}^{\prime}_{2},t)}{\int^{\prime}d^{\theta_{1}}\bm{r}^{\prime}_{1}d^{\theta_{2}}\bm{r}^{\prime}_{2}},
\end{equation}
and also define $\mu_{(2m-\theta)}(t)$ as~\cite{footnote1}
\begin{equation}
\mu_{(2m-\theta)}(t) \equiv \int^{\prime}d^{\theta_{1}}\bm{r}_{1}d^{\theta_{2}}\bm{r}_{2}\eta(\bm{r}_{1},\bm{r}_{2},t). 
\end{equation} 

Replacing the integral over the propagator in Eq.~(\ref{defectmt}) with $\widetilde{\nu}_{(2m-\theta)}(\epsilon)$ yields
\begin{equation}
\int^{\prime}d^{\theta_{1}}\bm{r}_{1}d^{\theta_{2}}\bm{r}_{2}\widetilde{P}(\bm{r}_{1},\bm{r}_{2},\epsilon) = \frac{\widetilde{\mu}_{(2m-\theta)}(\epsilon)}{1+\mathcal{C}_{(2m-\theta)}\widetilde{\nu}_{(2m-\theta)}(\epsilon)},
\end{equation}
which is an approximate result. Consequently, substituting this result into Eq.~(\ref{Itildemt1}) gives
\begin{equation}
\label{Itildemt2}
\widetilde{\mathcal{I}}(\epsilon) =\frac{1}{\epsilon}\left[\frac{\widetilde{\mu}_{(2m-\theta)}(\epsilon)}{1/\mathcal{C}_{(2m-\theta)}+\widetilde{\nu}_{(2m-\theta)}(\epsilon)}\right].
\end{equation}
This is the generalized form of the infection probability (in the Laplace domain) given in Eq.~(\ref{Itildemm2}), to cases with differing dimensionality between the infection region and motion of the walkers. The effective rate of infection is given by
\begin{equation}
\alpha = \frac{\mu_{(2m-\theta)}(\infty)}{1/\mathcal{C}_{(2m-\theta)}+1/\mathcal{M}_{(2m-\theta)}},
\end{equation}
with
\begin{equation}
\frac{1}{\mathcal{M}_{(2m-\theta)}} = \int_{0}^{\infty}dt\,\left[\nu_{(2m-\theta)}(t)-\mu_{(2m-\theta)}(t)\right].
\end{equation}

We proceed to derive $\nu_{(2m-\theta)}$ and $\mu_{(2m-\theta)}$ in CM-relative coordinates. This leads to
\begin{equation}
\label{nuPM}
\nu_{(2m-\theta)}(t)\equiv \frac{\int^{\prime}d^{\theta_{+}}\bm{r}^{\prime}_{+}d^{\theta_{-}}\bm{r}^{\prime}_{-}\int^{\prime}d^{\theta_{+}}\bm{r}_{+}d^{\theta_{-}}\bm{r}_{-}\Pi(\bm{r}_{+},\bm{r}^{\prime}_{+},\bm{r}^{\prime}_{-},\bm{r}^{\prime}_{-}t)}{\int^{\prime}d^{\theta_{+}}\bm{r}^{\prime}_{+}d^{\theta_{-}}\bm{r}^{\prime}_{-}},
\end{equation}
and
\begin{equation}
\mu_{(2m-\theta)}(t) \equiv \int^{\prime}d^{\theta_{+}}\bm{r}_{+}d^{\theta_{-}}\bm{r}_{-}\eta(\bm{r}_{+},\bm{r}_{-},t).
\end{equation} 

It was shown in the last section that the geometry of the infection region does not depend on the CM coordinate when we consider for infection transmission to take place as a function of the relative distance of walkers. This means that $\theta_{+}=m$ while $\theta_{-}$ remains to be determined for a specific scenario at hand. Thus, the primed integral is written as
\begin{equation}
\int^{\prime}d^{\theta_{+}}\bm{r}_{+}d^{\theta_{-}}\bm{r}_{-} \rightarrow \int_{\text{all space}}d^{m}\bm{r}_{+}\int^{\prime}d^{\theta_{-}}\bm{r}_{-},
\end{equation}
and $2m-\theta = 2m-\theta_{+}-\theta_{-} = m-\theta_{-}$.
Performing the first integral in the numerator of the definition of $\nu_{(2m-\theta)}(t)$ given in Eq.~(\ref{nuPM}) results in 
\begin{equation}
\int_{\text{all space}}d^{m}\bm{r}_{+}\int^{\prime}d^{\theta_{-}}\bm{r}_{-}\Pi(\bm{r}_{+},\bm{r}_{+}^{\prime},\bm{r}_{-},\bm{r}_{-}^{\prime},t) = \int^{\prime}d^{\theta_{-}}\bm{r}_{-}\,\Pi(\bm{r}_{-},\bm{r'}_{-},t), 
\end{equation}
where $\Pi(\bm{r}_{-},\bm{r'}_{-},t)$ is given in Eq.~(\ref{pirel}). With this result, the $\nu$-function can be written as
\begin{equation}
\label{mainPMnu}
\nu_{(2m-\theta)}(t) = \nu_{(m-\theta_{-})}(t) = \frac{\int^{\prime}d^{\theta_{-}}\bm{r}^{\prime}_{-}\int^{\prime}d^{\theta_{-}}\bm{r}_{-}\Pi(\bm{r}_{-},\bm{r}^{\prime}_{-},t)}{\int^{\prime}d^{\theta_{-}}\bm{r}^{\prime}_{-}}.
\end{equation}
For a $\delta$-function initial condition, $\mu_{(2m-\theta)}(t)$ becomes
\begin{equation}
\label{mainPMmu}
\mu_{(2m-\theta)}(t)=\mu_{(m-\theta_{-})}(t) = \int^{\prime}d^{\theta_{-}}\bm{r}_{-}\Pi(\bm{r}_{-},\bm{r}^{0}_{-},t).
\end{equation}

\subsection{Specific Example of a Reaction Region as a Ring}
We present here the simplest example in 2-dimensions, where the transmission of infection is assumed to occur when the mice are \emph{at} a certain distance from each other.  We call this distance the infection range. In polar coordinates, such geometry of the infection region becomes a ring of a given radius $b$, with $b^2 = \left(x_{-}^{1}\right)^2+\left(x_{-}^{2}\right)^2$. The dimensionality of the ring infection-region in the relative coordinate is $1$. Thus, the subscript $(2m-\theta)$ becomes $(2m-\theta) = m-\theta_{-}=2-1=1$.\\

Note that, for a ring infection-region of radius $b$, the primed integral is explicitly written as
\begin{equation}
\int^{\prime}d^{1}\bm{r}_{-} \rightarrow \int_{0}^{\infty}dr\int_{0}^{2\pi}r d\phi\,\frac{1}{r 2\pi}\delta\left(r-b\right) = \int_{0}^{\infty}dr\int_{0}^{2\pi}d\phi\,\frac{1}{2\pi}\delta\left(r-b\right),
\end{equation}
in the polar coordinates. We calculate $\nu_{1}(t)$ and $\mu_{1}(t)$ from the definitions given in Eq's.~(\ref{mainPMnu}) and (\ref{mainPMmu}). 
\begin{equation}
\label{nuRing}
\nu_{1}(t) = \frac{\int_{0}^{\infty}dr'\int_{0}^{2\pi}d\phi'\,\frac{1}{2\pi}\delta\left(r'-b\right)\int_{0}^{\infty}dr\int_{0}^{2\pi}d\phi\,\frac{1}{2\pi}\delta\left(r-b\right)\,\Pi(r,r',\phi,\phi',t)}{\int_{0}^{\infty}dr'\int_{0}^{2\pi}d\phi'\,\frac{1}{2\pi}\delta\left(r'-b\right)} = \frac{1}{2\pi}\int_{0}^{2\pi}d\phi'\,\Pi(b,b,\phi',t),
\end{equation}
and 
\begin{equation}
\mu_{1}(t) = \int_{0}^{\infty}dr\int_{0}^{2\pi}d\phi\,\frac{1}{2\pi}\delta\left(r-b\right)\,\Pi(r,r^{0},\phi,\phi^{0},t) = \frac{1}{2\pi}\Pi(b,r^{0},\phi^{0},t).
\end{equation}
Here $\Pi(r,r',\phi,\phi',t)$ and $\Pi(r,r',\phi',t)$ are given in Eq's.~(\ref{PiPolar1}) and (\ref{PiPolar2}), respectively.\\

The ring infection-region in 2-dimension is analogous to a point infection-region in 1-dimension in the following way. The analysis provided at the beginning of this section shows that the $\nu$-function for the point infection in 1-dimension behaves as $t^{-1/2}$ for small times. It is straightforward to demonstrate $\nu_{1}(t)$ given in Eq.~(\ref{nuRing}) exhibits the same time-dependence for small times. As $t\rightarrow0$, the argument of the Bessel function becomes large where its asymptotic form is appropriate, as given in Eq.~(\ref{I0tShort}), and $\nu_{1}(t)$ behaves as, at short times,
\begin{equation}
\nu(t\rightarrow0) \rightarrow \frac{1}{2\pi}\int_{0}^{2\pi}d\theta'\left(\frac{1}{4\pi D t}\right)e^{-\frac{b^2}{2Dt}}\sqrt{\frac{2Dt}{b^2}}e^{\frac{b^2}{2Dt}}\sim t^{-1/2}.
\end{equation}

\subsubsection{Recovery of Nonmonotonicity and Appearance of a Resonance Effect}

\begin{figure}[h!] % float placement: (h)ere, page (t)op, page (b)ottom, other (p)age
  \centering
   \includegraphics[width= 0.8\columnwidth]{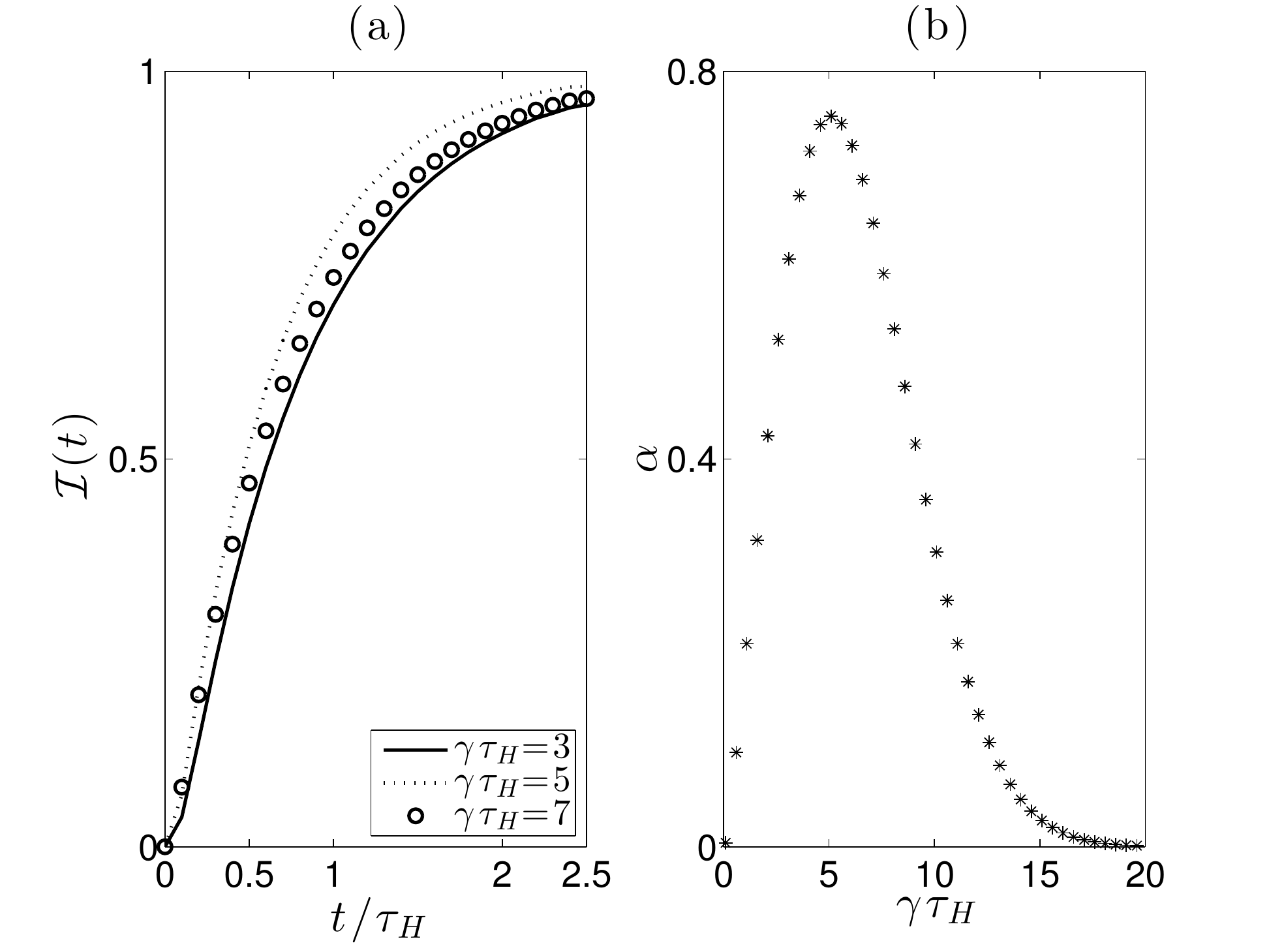}
   \caption{(a) The non-monotonic effect in the infection probability $\mathcal{I}(t)$ as a function of increasing strength of confinement, $\gamma$. Here $\gamma$ is scaled to $1/\tau_H$, the diffusive time to for the walker to traverse the inter-home distance $H$. The infection probability $\mathcal{I}(t)$ is plotted against time in units of $\tau_H$. Each infection curve corresponds to a different value of $\gamma\tau_H$. (b) The effective rate $\alpha$ in units of $\tau_H$ is plotted against $\gamma\tau_H$. The peak indicates the optimal value of $\gamma\tau_H$ for transmission of infection. The rate of infection $\mathcal{C}_{1}$ is illustrative and kept constant at 9 in units of 2D/H.}
   \label{fig:SK1_nonMon}
\end{figure} 

To obtain the infection probability, the integral over $\phi'$ of  $\nu_{1}(t)$ given in Eq.~(\ref{nuRing}) is done numerically. Then this result and $\mu_{1}(t)$ of Eq.~(\ref{muDisk}) are numerically Laplace-transformed and substituted into Eq.~(\ref{Itildemt2}) to obtain $\widetilde{\mathcal{I}}(\epsilon)$. Finally, $\widetilde{\mathcal{I}}(\epsilon)$ is numerically Laplace inverted to obtain the result, $\mathcal{I}(t)$, in the time domain.\\

The resulting infection curve and the effective rate showing the non-monotonic behavior due to the effect of confinement are shown in Fig.~\ref{fig:SK1_nonMon}. In panel (a), the infection curve is plotted against time. The latter is scaled to the diffusive time $\tau_H=H^2/2D$. Each curve corresponds to a value of the unitless parameter $\gamma\tau_H$, effectively to a given potential strength $\gamma$. Panel (b) of Fig.~\ref{fig:SK1_nonMon} shows the effective rate $\alpha$ scaled to $1/\tau_H$, plotted against $\gamma\tau_H$ again showing the non-monotonic effect. Intuitively, one can think of this effect arising from the changing probability of the walkers finding one another given their inter-home distance $H$ and the degree of their confinement to respective homes, in the contact-limited case. The steady-state Smoluchowski width, $\sigma$ essentially gives the degree of confinement. When these lengths are balanced, \emph{i.e.}, $H=\sigma$, the probability for the walkers to find each other is maximized, which in tern is reflected in the maximized infection probability.\\

The infection range, $b$ is a new length scale in addition to $H$ and $\sigma$. It arises from the extension of the infection influence to a finite region and enters the dynamics in such a manner to effectively modulate the role of $H$ per given $\sigma$. In other words, when the walkers live at homes that are separated by $H$ and if the the infection happens at a range $b$, then part of the problem becomes equivalent to a co-location infection with $|H-b|$ as the home separation distance. For this reason, when $H=b$, \emph{i.e.}, when the inter-home distance and infection range are the same, infection occurs as if though the walkers share a home. In such a situation, increasing the strength of confinement only helps the infection transmission. We refer to this phenomenon as `resonant' behavior.\\

\begin{figure}[h!] % float placement: (h)ere, page (t)op, page (b)ottom, other (p)age
  \centering
   \includegraphics[width= 0.8\columnwidth]{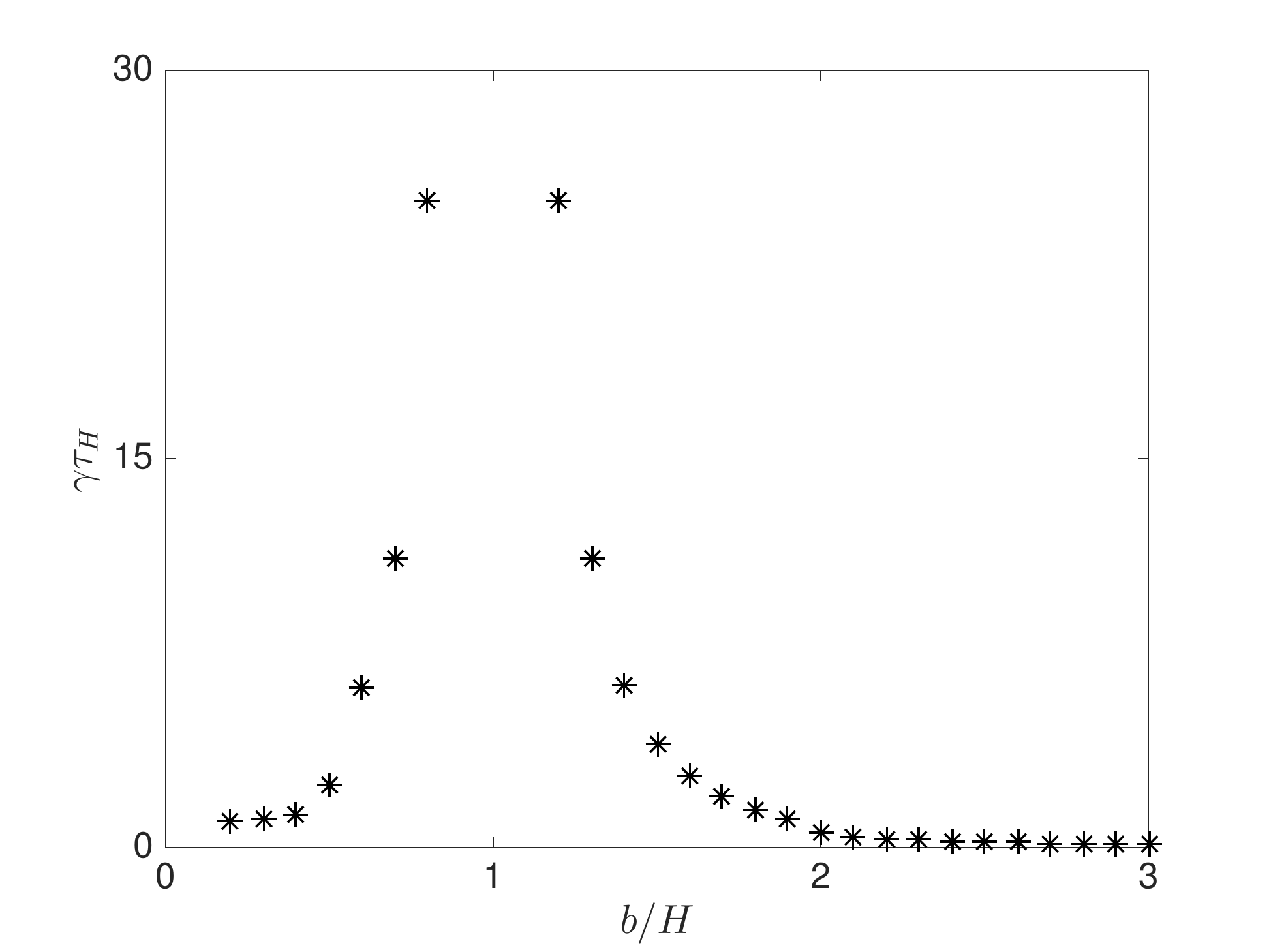}
   \caption{Optimal value of the confining potential steepness $\gamma$ (scaled to $1/\tau_H$) for maximum infection rate as a function of ($b/H$). The divergence of $\gamma\tau_H$ at $b/H=1$ signals the \emph{resonant} phenomenon. Values given here are accurate for a large value of the infection rate.}
   \label{fig:1DBoxer_gammaTauAH}
\end{figure}
 
We show this effect in Fig.~\ref{fig:1DBoxer_gammaTauAH}  for the simplest case where 
an infection range is introduced in 1-dimension. The optimal value of $\gamma$ (scaled to $1/\tau_H$)  for infection transmission is plotted as a function of the ratio between the infection range and inter-home distance, $b/H$. The resonant behavior is indicated by the diverging value of $\gamma\tau_{H}$ at $b/H =1$. We provide a detailed analysis to determine the optimal value of $\gamma\tau_H$ in Appendix~\ref{1DBoxerAP}.\\ 

\section{Discussion}
The analysis of transmission of infection in an epidemic such as the Hantavirus which propagates by contact was begun in ref.~\cite{KSBMB14} and is completed in the present paper. The essential theory is in ref.~\cite{KSBMB14} and the extension to higher dimensions is in the present paper. Produced in~\cite{KSBMB14} is a reaction-diffusion formulation of two mice as Smoluchowski random walkers transmitting infection to one another upon encounter at a point. While the theory poses no problem in 1-dimension, one finds a singularity in the behavior of $\nu$-function at short times in higher dimensions.\\

This peculiarity occurs due to  a mathematical problem of the vanishing probability to find a point in dimensions higher than one (in continuum). A clear way to remove the singularity, hence to recover the existence of the Laplace transform of the $\nu$-function, is to consider an infection region to have a finite extent. We have presented such an extension here. It includes a finite infection region. We demonstrate its use explicitly in 2-dimensions via two types of infection region. One corresponds to a case when the infection is transmitted when the mice are \emph{within} a certain distance of each other, and the other is the case when they are \emph{at} a certain distance from one another. Respectively, the region becomes a disk and a ring of certain radius in the polar coordinate representation of the problem. It was shown that, as fully expected, the non-monotonic effect of confinement on the infection probability generalizes to higher dimensions. However we have also discovered new effect. When the infection distance and inter-home distance are comparable, a resonant phenomenon is observed in the optimal value of the confinement for the infection transmission to occur. This is shown explicitly for the ring infection region case.\\

%%%%%%%%%%%%%%%%%%%%%%%%%%%%%%
\appendix
%%%%%%%%%%%%%%%%%%%%%%%%%%%%%%
\counterwithin{figure}{section}
%Appendix: Satomi's verification of known results (without trapping) to gain confidence.
\section{Analysis in 1-dimension: Introduction of Infection Range}
\label{1DBoxerAP}
We consider that the infection is transmitted \emph{at} a range $b$ while the mice move in 1-dimension, and start the analysis in CM-relative coordinates. The equation of motion for $P(x_{+},x_{-},t)$, the probability density to find the walkers at CM and relative coordinates $x_+$ and $x_-$, respectively, at time $t$ is 
\begin{align}
\frac{\partial P(x_+,x_-,t)}{\partial t} &= \frac{\partial}{\partial x_{+}}\gamma\left(x_{+}-h_{+}\right)P(x_+,x_-,t)+\frac{\partial}{\partial x_{-}}\gamma\left(x_{-}-h_{-}\right)P(x_+,x_-,t) \nonumber \\
&\quad\quad +D\left(\frac{\partial^2}{\partial x_{+}^2}+\frac{\partial^2}{\partial x_{-}^2}\right) P(x_+,x_-,t)\nonumber \\
&\quad\quad -\mathcal{C}_1\delta\left(x_{-}-b\right)P(x_{+},x_{-},t)-\mathcal{C}_{1}\delta\left(x_{-}+b\right)P(x_{+},x_{-},t). \label{Boxer1DPEQ}
\end{align}
The first three terms represent the Smoluchowski motion. The infection transmission is described in the last two terms; the arguments of the $\delta$-functions indicate that the infection is transmitted to the susceptible mouse from the infected one when they are a distance $b$ apart, \emph{i.e.}, when $x_{-}=\pm b$. The infection rate is given by $\mathcal{C}_{1}$.\\

The symmetry and simplicity of this infection region allows for an exact calculation of $\tilde{\nu}(\epsilon)$, without the approximation of the $\nu$-function method. The propagator for this problem is given by
\begin{equation}
\Pi(x_{+},x'_{+},x_{-},x'_{-},t) = \frac{1}{4\pi D\mathcal{T}(t)}e^{-\frac{\left(x_{+}-h_{+}-(x'_{+}-h_{+})e^{-\gamma t}\right)^2+\left(x_{-}-h_{-}-(x'_{-}-h_{-})e^{-\gamma t}\right)^2}{4D\mathcal{T}(t)}}.
\end{equation}
In terms of this propagator, the solution to Eq.~(\ref{Boxer1DPEQ}) in the Laplace domain as
\begin{align}
\tilde{P}(x_{+},x_{-},\epsilon) &= \tilde{\eta}(x_{+},x_{-},\epsilon) \nonumber \\
& -\mathcal{C}_{1}\int_{-\infty}^{\infty}dx'_{+}\tilde{\Pi}(x_{+},x'_{+},x_{-},-b,\epsilon)\tilde{P}(x'_{+},-b,\epsilon) \nonumber \\
& -\mathcal{C}_{1}\int_{-\infty}^{\infty}dx'_{+}\tilde{\Pi}(x_{+},x'_{+},x_{-},b,\epsilon)\tilde{P}(x'_{+},b,\epsilon).
\label{PseudoSolBoxer1D}
\end{align}
From its definition given in Eq.~(\ref{Itildemm1}), the infection probability in the Laplace domain is given via Eq.~(\ref{PseudoSolBoxer1D}) above by
\begin{equation}
\widetilde{\mathcal{I}}(\epsilon) = \frac{\mathcal{C}_{1}}{\epsilon}\left[\int_{-\infty}^{\infty}dx_{+}\widetilde{P}(x_{+},b,\epsilon)+\int_{-\infty}^{\infty}dx_{+}\widetilde{P}(x_{+},-b,\epsilon)\right].
\end{equation}

Use of the defect technique, \emph{i.e.}, setting $x_{-}=\pm b$ and integrating $x_{+}$ over all space, after some algebra, yields
\begin{align}
&\int_{-\infty}^{\infty}dx_{+}\widetilde{P}(x_{+},b,\epsilon)+\int_{-\infty}^{\infty}dx_{+}\widetilde{P}(x_{+},-b,\epsilon)\nonumber \\
& = \frac{1}{\mathcal{C}_{1}}\frac{\left[1/\mathcal{C}_{1}+\tilde{\nu}^{++}_{1}(\epsilon)-\tilde{\nu}^{+-}_{1}(\epsilon)\right]\tilde{\mu}^{-}_{1}(\epsilon)+\left[1/\mathcal{C}_{1}+\tilde{\nu}^{--}_{1}(\epsilon)-\tilde{\nu}^{-+}_{1}(\epsilon)\right]\tilde{\mu}^{+}_{1}(\epsilon)}{\left[1/\mathcal{C}_{1}+\tilde{\nu}^{--}_{1}(\epsilon)\right]\left[1/\mathcal{C}_{1}+\tilde{\nu}^{++}_{1}(\epsilon)\right]-\tilde{\nu}^{-+}_{1}(\epsilon)\tilde{\nu}^{+-}_{1}(\epsilon)},
\end{align} 
where
\begin{align}
\mu^{+}_{1}(t) &=\int_{-\infty}^{\infty}dx_{+}\eta(x_{+},b,t) = \frac{1}{\sqrt{4\pi D\mathcal{T}(t)}}\,e^{-\frac{\left(b-H-(x_{-}^{0}-H)e^{-\gamma t}\right)^2}{4D\mathcal{T}(t)}}\\
\mu^{-}_{1}(t) &=\int_{-\infty}^{\infty}dx_{+}\eta(x_{+},-b,t) = \frac{1}{\sqrt{4\pi D\mathcal{T}(t)}}\,e^{-\frac{\left(b+H+(x_{-}^{0}-H)e^{-\gamma t}\right)^2}{4D\mathcal{T}(t)}}\\
\nu^{++}_{1}(t) & = \int_{-\infty}^{\infty}dx_{+}\Pi(x_{+},b,x'_{+},b)= \frac{1}{\sqrt{4\pi D\mathcal{T}(t)}}\,e^{-\frac{\left[(b-H)(1-e^{-\gamma t})\right]^2}{4D\mathcal{T}(t)}}\\
\nu^{+-}_{1}(t) & = \int_{-\infty}^{\infty}dx_{+}\Pi(x_{+},b,x'_{+},-b)=\frac{1}{\sqrt{4\pi D\mathcal{T}(t)}}\,e^{-\frac{\left[b-H+(b+H)e^{-\gamma t}\right]^2}{4D\mathcal{T}(t)}}\\
\nu^{-+}_{1}(t) & =\int_{-\infty}^{\infty}dx_{+}\Pi(x_{+},-b,x'_{+},b)= \frac{1}{\sqrt{4\pi D\mathcal{T}(t)}}\,e^{-\frac{\left[b+H+(b-H)e^{-\gamma t}\right]^2}{4D\mathcal{T}(t)}}\\
\nu^{--}_{1}(t) & =\int_{-\infty}^{\infty}dx_{+}\Pi(x_{+},-b,x'_{+},-b)= \frac{1}{\sqrt{4\pi D\mathcal{T}(t)}}\,e^{-\frac{\left[(b+H)(1-e^{-\gamma t})\right]^2}{4D\mathcal{T}(t)}}.
\end{align}
In the calculation of the $\mu_{1}(t)$'s, a $\delta$-function initial condition was assumed. The infection probability in the Laplace domain is then given exactly by 
\begin{equation}
\widetilde{\mathcal{I}}(\epsilon) = \frac{1}{\epsilon}\frac{\left[1/\mathcal{C}_{1}+\tilde{\nu}^{++}_{1}(\epsilon)-\tilde{\nu}^{+-}_{1}(\epsilon)\right]\tilde{\mu}^{-}_{1}(\epsilon)+\left[1/\mathcal{C}_{1}+\tilde{\nu}^{--}_{1}(\epsilon)-\tilde{\nu}^{-+}_{1}(\epsilon)\right]\tilde{\mu}^{+}_{1}(\epsilon)}{\left[1/\mathcal{C}_{1}+\tilde{\nu}^{--}_{1}(\epsilon)\right]\left[1/\mathcal{C}_{1}+\tilde{\nu}^{++}_{1}(\epsilon)\right]-\tilde{\nu}^{-+}_{1}(\epsilon)\tilde{\nu}^{+-}_{1}(\epsilon)}.
\end{equation}
The nonmonotonic effect was explained in detail ref.~\cite{KSBMB14} in the steady state and in the contact limit where $1/\mathcal{C}_{1}$ is much greater than any of the $\nu_{1}(t)$'s and $\mu_{1}(t)$'s in its effect. In these lilmit, $\mathcal{I}(t)$ approximately becomes
\begin{equation}
\label{ItContactLim1DBoxer}
\mathcal{I}(t) \sim \mathcal{C}_{1}\left(\mu^{+}_{1}(\infty)+\mu^{-}_{1}(\infty)\right)\cdot t, 
\end{equation} 
where we note that $\nu^{++}_{1}(\infty)=\nu^{+-}_{1}(\infty)=\mu^{+}_{1}(\infty)$ and $\nu^{--}_{1}(\infty)=\nu^{-+}_{1}(\infty)=\mu^{-}_{1}(\infty)$. The condition for the optimal value of $\gamma\tau_H$ value found from this result yields the transcendental relation,
\begin{equation}
\frac{1-2\gamma\tau_H(1-b/H)^2}{1-2\gamma\tau_H(1+b/H)^2} = -e^{-4\gamma\tau_H(b/H)}.
\end{equation}
The value of $\gamma\tau_H$ for given value of $b/H$ found from this equation is plotted in Fig.~\ref{fig:1DBoxer_gammaTauAH}.


\begin{thebibliography}{99}
\bibitem{footnote1}
Note that when $(2m-\theta) =0$ the $\nu$-function is denoted simply by $\nu(t)$ (the original definition~\cite{vmknu}). In a couple of recent publications~\cite{KSinJPC,KSBMB14}, the symbol $\nu(t)$ has been applied to a function with varying dimensions according to the dimension of the motion and the trap.
\end{thebibliography}
\end{document}